\documentclass[12pt]{article}
\usepackage[dvips]{graphicx}
\usepackage{pdproc}

\newcommand{\lesssim}{ \mathop{}_{\textstyle \sim}^{\textstyle <} }

\makeatletter 
\def\@cite#1{[#1]} 
\makeatother    
\begin{document}

\renewcommand{\thefootnote}{\alph{footnote}}

\title{
  New Constraint on Squark Flavor Mixing from $^{199}$Hg EDM~\footnote{
    To appear in the Proceedings of {\it SUSY2004}, Tsukuba, Japan, June 17-23, 
    2004}
}

\author{ MOTOI ENDO }

\address{ Department of Physics, Tohoku University,
Sendai 980-8578, Japan
%%%%% You may comment out the e-mail address line below.  
\\ {\rm E-mail: endo@tuhep.phys.tohoku.ac.jp}}

\abstract{
  We obtain a new constraint on the CP-violating flavor mixing 
  between the left-handed top scalar quark $(\tilde{t}_L)$ and charm 
  scalar quark $(\tilde{c}_L)$, by considering a chargino loop 
  contribution to chromo-electric dipole moment of strange quark, 
  which is limited by the electric dipole moment of the neutron and atoms.  
  It is found that the flavor mixing should be suppressed to 
  the level of ${\cal O}(0.1)$ for the CP phase of order unity.  
  Although it is much stronger than the known constraint from 
  the chargino loop contribution to $b \to s \gamma$, the moderate 
  constraint we obtain here is argued to leave room for sizable 
  supersymmetric contribution to the CP asymmetry in $B_d^0 \to \phi K^0$.
}

\normalsize\baselineskip=15pt

\section{Introduction}

Suppression of the flavor changing neutral current (FCNC) by 
the GIM mechanism is one of the most celebrated features of the 
Standard Model (SM).  This mechanism is applied because of either 
degeneracy of the masses or small generation mixings.  On the contrary, 
the SUSY models do not generally satisfy these conditions.  
Thus they tend to suffer from large FCNC.  Actually, $1-2$ and $1-3$ 
generation mixings in the squark mass matrices are stringently constrained 
by the $K^0-\bar K^0$ mixing and so on.  
On the other hand, $2-3$ squark mixings are 
allowed to be relatively large by the known bound from 
${\rm Br}(b \to s \gamma)$, which provides $(\delta_{LL,RR}^{d})_{23} \lesssim 
O(10^{-1})-O(1),~(\delta_{LL}^{u})_{23} \lesssim O(1) - O(10)$.  
Therefore we expect to detect contributions from the SUSY models 
by observing the transition processes of $2-3$ generation.  

Nowadays a special attention is paid to the $b \to s$ flavor mixings 
in B mesons.  A very interesting process is the mixing-induced CP asymmetry 
of $B_d \to \phi K^0$, $S_{\phi K^0}$.  Experimentally, the latest result of 
the Belle collaboration announced the deviation from the SM that~\cite{Belle}, 
\begin{eqnarray}
  S_{\phi K^0}(\mbox{Belle}) \ =\ 0.06 \pm 0.33 \pm 0.09,
\end{eqnarray}
while that of the BaBar collaboration~\cite{Aubert:2004dy},
\begin{eqnarray}
  S_{\phi K^0}(\mbox{BaBar}) \ =\ 0.50 \pm 0.25 ^{+0.07}_{-0.04},
\end{eqnarray}
is consistent with the SM prediction. Though the situation is yet unclear, 
one certainly has to watch what is happening when more data are accumulated, 
and on the theoretical side it is important to investigate in what situation 
new physics can generate large $b \to s$ transition.  

In SUSY models, a new source of flavor mixing originates in SUSY breaking 
masses of squarks and sleptons.  It is convenient to parameterize these squark 
mixing as
\begin{eqnarray}
  (\delta^u_{LL})_{ij}=\frac{(m^2_{\tilde u_L})_{ij}}{m^2_{\tilde q}}, 
  \quad
  (\delta^u_{RR})_{ij}=\frac{(m^2_{\tilde u_R})_{ij}}{m^2_{\tilde q}},
\end{eqnarray}
where $(m^2_{\tilde u_{L(R)}})_{ij}$ is an $(ij)$ element of the left-handed 
(right-handed) squark mass squared matrix for up-type squarks in the superCKM 
basis.  A similar notation will be used for the down-type squarks.  
It is important that there is chirality structure in the squark mass 
matrices.  The flavor mixings between the second and third generations 
both in left-handed squarks (LL) and in right-handed squarks (RR) have 
CP-violating phase generally, and thus both/either of the mixings can 
make contribution to $S_{\phi K^0}$.  

A notable observation has recently been made in Ref.~\cite{Hisano:2003iw}, 
which has found that the present experimental bound on the electric 
dipole moment (EDM) of the neutron and atoms including the mercury 
severely constrains the CP violating part of the squark mixings.  
In fact, the diagrams which have two sources of $2-3$ generation 
mixings contribute to the strange-quark color electric dipole moment 
(CEDM) generally.  Specifically they consider a product of 
$(\delta^d_{LL})_{23}$ and $(\delta^d_{RR})_{32}$.  The reason is that 
in the presence of sizable $\tilde{b}_L-\tilde{s}_L$ and 
$\tilde{b}_R-\tilde{s}_R$ mixings, the dominant contribution to 
the CEDM of the strange quark, $d_s^C$, arises from the gluino-loop 
diagram with the $\tilde{b}_L-\tilde{b}_R$ mixing, which is proportional 
to $m_b \mu \tan \beta$, and thus is enhanced by $m_b/m_s$ over 
the usual contribution induced by the $\tilde{s}_L-\tilde{s}_R$ mixing.  
On the other hand, the strange CEDM is bounded by the null measurement 
of the neutron and atomic EDMs.  Recently the contributions of strange 
CEDM to the EDMs are revisited and the neutron is shown to provide the most 
stringent constraint, rather than the mercury~\cite{Hisano:2004tf}.  
As a result, the CP-violating part of the product of the squark mixings 
are bounded very strongly by the neutron EDM.  

Although the constraint is only for the product of $(\delta^d_{LL})_{23}$ 
and $(\delta^d_{RR})_{32}$, radiative corrections (renormalization group 
effects) due to Yukawa interaction for up-type quark masses generate 
significant flavor mixing in the LL sector when one considers the high 
scale SUSY breaking scenario where the mediation of the SUSY breaking 
takes place at a high energy scale.  Given the non-negligible mixing 
in the LL sector at the electroweak scale, the bound from the hadronic 
EDMs thus practically constrains the flavor mixing in the RR sector.  
It has been shown that with the parameters satisfying the constraint, 
the contribution to the mixing-induced CP asymmetry of the 
$B_d \to \phi K^0$ from the RR mixing should be negligibly small.  
Thus we focus on the LL squark mixing.  

In this talk, which is based on the works with Mitsuru Kakizaki and 
Masahiro Yamaguchi in Ref.~\cite{Endo:2003te,Endo:2004xt}, we show that the 
hadronic EDMs provide a new constraint on the LL squark mixings between 
$2-3$ generations, through the chargino mediated diagrams.  
We also revisit the SUSY contributions to the mixing-induced CP asymmetry 
of $B_d \to \phi K^0$, considering the experimental bound from 
the hadronic EDMs as well as that from ${\rm Br}(b \to s \gamma)$.  

\section{LL Squark Mixing}

Here we point out that the LL mixing is constrained by the hadronic 
EDMs.  We first emphasize that chargino-mediated processes can also 
provide large contributions to the CEDM of the strange quark $d_s^C$
because there are diagrams which are enhanced by the top Yukawa coupling 
constant.  In the presence of $2-3$ mixing in the LL sector,
$d^C_s$ is induced by the double mass insertion diagram and is evaluated as
\begin{eqnarray}
  d_s^C &\simeq& \frac{1}{8 \pi^2} \frac{G_F}{\sqrt{2}}  
  m_t^2 m_s V_{ts}   \frac{|A_t \mu| \tan \beta}{m_{\tilde{q}}^4} 
  M_a(x) |(\delta_{LL}^u)_{32}| \sin \theta_a,
  \label{eq:cedm}
\\
  &\simeq& 2.8 \times 10^{-24}\ e\ \mbox{cm} 
  \times |(\delta_{LL}^u)_{32} \sin \theta_a |
  \nonumber \\
  && \times 
  \left( \frac{\tan \beta}{20} \right)
  \left( \frac{|\mu|}{250~\mbox{GeV}} \right)
  \left( \frac{|A_t|}{500~\mbox{GeV}} \right)
  \left( \frac{m_{\tilde{q}}}{500~\mbox{GeV}} \right)^{-4}
  \left( \frac{M_a(x)}{-0.31} \right).
\end{eqnarray}
where the loop function $M_a(x)$ is defined in Ref.~\cite{Endo:2003te} 
and $x \equiv {|\mu|^2}/{m_{\tilde{q}}^2}$.
Here $\theta_a$ parameterizes the CP violating phase as
$\theta_a = \arg [A_t \mu (\delta_{LL}^u)_{32}]$.
Note that one of the flavor mixings is now given by the CKM matrix.  
Thus compared with the experimental upper limit from the neutron EDM, 
we obtain
\begin{eqnarray}
  |(\delta_{LL}^u)_{32} \sin \theta_a| \ \lesssim\  0.07 (0.09),
\end{eqnarray}
from the neutron EDM.  Here the number in the parentheses represents 
the assumption of the Pecci-Quinn (PQ) symmetry.  
The relevant parameters are taken to be $\tan\beta = 20$, 
$|\mu| = 250\ {\rm GeV}$ and $|A_t| = m_{\tilde{q}} = 500\ {\rm GeV}$.
Therefore we conclude that the $\tilde{t}_L-\tilde{c}_L$ mixing angle 
or the CP violating phase must be suppressed at the level of $O(0.1)$ 
when $\tan\beta$ is large, in the light of the result of the neutron 
EDM experiment.  We stress that this constraint is severer than that 
obtained from the branching ratio of the process $b \to s \gamma$, 
by about one or two orders of magnitude depending on mass spectra of 
superparticles.  

\section{B Physics}

Finally we investigate the implication on $b \to s$ transition processes 
of B mesons from the constraints by the hadronic EDMs as well as 
${\rm Br}(b \to s \gamma)$.  Since the RR squark mixing is suppressed 
sufficiently, we consider the contributions from the LL mixing.  
The SUSY contributions to the processes are dominated by gluino mediated 
diagrams and thus we need the experimental constraint on the down-type squark 
mixing.  Indeed the bound from the hadronic EDMs obtained above is on the 
up-type one, but notice that $(\delta^u_{LL})_{ij}$ and $(\delta^d_{LL})_{ij}$ 
are related to each other because of the $SU(2)_L$ symmetry.  In fact, in the 
superCKM basis, one finds that
\begin{equation}
  (\delta^d_{LL})_{32} \sim (\delta^u_{LL})_{32} 
  + \lambda (\delta^u_{LL})_{31} + O(\lambda^2)
\end{equation}
in terms of the Wolfenstein parameter $\lambda \sim 0.2$.
Hence in the absence of accidental cancellation or hierarchy
among the parameters the bound on $(\delta^u_{LL})_{32}$ is translated into 
that on $(\delta^d_{LL})_{32}$.  

\begin{figure}[htb]
  \begin{center}
    \includegraphics*[scale=0.7]{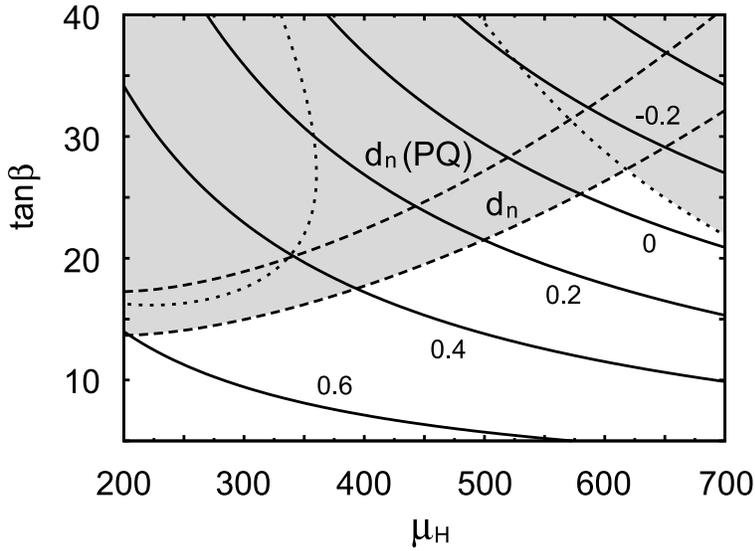}
    \caption{Constant contours of the CP asymmetry $S_{\phi K^0}$ (solid) 
      when the LL down-type squark mixing is $|(d_{LL}^d)_{23}|=0.1$ with 
      maximal phase.  The contours of the $b \to s \gamma$ branching ratio 
      (dotted, in units of $10^{-4}$) and the CEDM of the strange quark 
      which is constrained from the neutron EDM (dashed) are also shown.  
      The soft parameters are 
      $m_{\tilde{g}}^2 = m_{\tilde{q}}^2 = (500\ {\rm GeV})^2$ and 
      $A_t = 500\ {\rm GeV}$.}
    \label{fig:d01}
  \end{center}
\end{figure}

Let us now discuss the mixing-induced CP asymmetry of $B_d^0 \to \phi K^0$, 
which is known to be one of the most interesting modes. In Fig.~\ref{fig:d01}, 
we show the numerical result, where the constant contours of $S_{\phi K^0}$ 
are shown.  The constraints from the $b \to s \gamma$ branching ratio, 
for which we take a rather conservative bound, $2.0 \times 10^{-4} < 
{\rm Br}(b \to s \gamma) < 4.5 \times 10^{-4}$, and from the hadronic EDMs, 
that is, the neutron EDM with/without the assumption of the PQ symmetry, 
are also displayed on the graph.  Here we show the result of 
$|(\delta_{LL}^d)_{23}| = 0.1$ with maximal complex phase.  
As a reference, we also fix the soft parameters as $m_{\tilde{g}}^2 = 
m_{\tilde{q}}^2 = (500\ {\rm GeV})^2$ and $A_t = 500\ {\rm GeV}$.  
And we take the other model parameters: the Wino mass $M_2 = 250\ {\rm GeV}$ 
and the Higgs mass parameters $m_{Hd}^2 = -m_{Hu}^2 = (250\ {\rm GeV})^2$.  
We find that the CP asymmetry becomes large as $\mu_H$ or $\tan\beta$ 
increases.  Consequently, from Fig. \ref{fig:d01}, we perceive that the 
contribution from the LL squark mixing to $B_d \to \phi K^0$ can become 
as large as $S_{\phi K} \lesssim 0$.  
%%%%%%%%%%%%%%%%%%%%%%%%%%%%%%%%
%%%%%%%%%%%%%%%%%%%%%%%%%%%%%%%%

\section{Acknowledgments}

The author thanks the Japan Society for the Promotion of Science for financial 
support.

\bibliographystyle{plain}

\end{document}